# X-ray Microdiffraction Images of Antiferromagnetic Domain Evolution in Chromium

P. G. Evans,[1*] E. D. Isaacs,[1] G. Aeppli,[2] Z. Cai,[3] B. Lai[3]

[1] Bell Laboratories, Lucent Technologies, 600-700 Mountain Ave., Murray Hill, NJ 07974.  [2] NEC Research Institute, 4 Independence Way, Princeton, NJ 08540.  [3] Advanced Photon Source, Argonne National Laboratory, 9700 S. Cass Ave., Argonne, IL 60439

[*] To whom correspondence should be addressed.  E-mail: pevans2@lucent.com

**Abstract**

Magnetic x-ray diffraction combined with x-ray focusing optics is used to image individual antiferromagnetic spin density wave domains in a chromium single crystal at the micron scale. The cross section for non-resonant magnetic x-ray scattering depends on the antiferromagnetic modulation vector and spin polarization direction and allows these quantities to be extracted independently. The technique is used to show that the broadening of the nominally first order "spin-flip" transition at 123 K, at which the spins rotate by 90°, originates at the walls between domains with orthogonal modulation vectors. During cooling the transition begins at these walls and progresses inwards. The modulation-vector domains are themselves unchanged.

The antiferromagnetic ordering of elemental chromium was guessed in 1936 by Louis Néel but had to await the invention of neutron diffraction for confirmation (*1*). Body-centered cubic Cr is the quintessential example of a structurally simple metal exhibiting complex magnetism. Electron spins order in a long-period density wave and on cooling not only undergo a transition between a disordered paramagnetic phase and an ordered Néel state, at $T_N$, but also display a transition between Néel states of mutually perpendicular spin polarizations, at $T_{SF}$. While this latter "spin flip" transition dominates many physical properties of chromium below room temperature, the mechanism and dynamics of nucleation and growth of antiferromagnetic domains remains elusive because no suitable microscopies have combined sensitivity to antiferromagnetism and spatial resolution (*2,3*). We describe here a magnetic x-ray diffraction technique exploiting focusing optics to visualize antiferromagnetic domains at the micron scale. As an example of this technique we show that the spin flip transition begins at the walls between domains with orthogonal propagation directions for the spin modulation and coarsens inward.

Coulomb interactions among electrons and the geometry of the Fermi surface within which electrons reside conspire (*4-6*) to produce the spin density wave (SDW) in Cr below the Néel transition at 311 K. The SDW has an amplitude, or peak magnetic moment, of 0.6 $\mu_B$ per atom and a modulation vector that can lie along any of the three <001> axes of the Cr crystal lattice (*7*). The transition we are interested in is not the Néel transition, but the spin flip transition near $T_{SF}$=123K, where the SDW changes between transverse and longitudinal polarization states (Fig. 1A). One signature of the spin flip transition occurs in the bulk magnetic susceptibility, which undergoes a small step at the critical temperature (Fig. 1B). Although magnetic susceptibility data allow the identification of anomalies, they do not by themselves provide their explanations. The most useful tools in this respect are magnetic neutron and X-ray diffraction, where scattering arises from electron spins in the same way as from atomic cores in conventional neutron and X-ray diffraction. Data are collected in reciprocal rather than direct space, and contain sharp Bragg peaks derived from coherent spin density modulations. For Cr, with lattice constant $a$, the period $Da$ of the spin modulation is incommensurate with the Cr lattice. The ordered magnetic state leads to new Bragg reflections at distances $\delta=1/D\approx0.05$ in reciprocal space from the points $(h,k,l)$ where $h+k+l$ is odd, at which Bragg peaks due to atomic cores are forbidden for the body centered cubic structure (*8,9*). The SDW also induces modulations in the charge and lattice that are proportional to the square of the spin density wave - this follows because the coupling between charge and spin is invariant with respect to spin reversal, allowing only even powers of the spin to couple to charge. The resulting charge density wave (CDW) then has half the period of the SDW, leading to additional non-magnetic Bragg peaks displaced by $2\delta$ from the $(h,k,l)$ points with even $h+k+l$ (*10*). The locations of SDW and CDW reflections near (001) and (002) in reciprocal space are shown (Fig. 1C).

For ferromagnets, magnetic domains are characterized by a single parameter, the spin polarization. For antiferromagnetic Cr, two parameters are needed – the spin polarization **S** and modulation wavevector **Q**, which can be independently determined via x-ray diffraction using the magnetic scattering reflection arising directly from the SDW and the conventional Thomson scattering from the CDW (*9*). To spatially resolve **S** and **Q**, magnetic x-ray microscopy experiments were performed using a hard x-ray microprobe facility at the Advanced Photon Source of Argonne National Laboratory. The source of synchrotron radiation was an undulator insertion device with the photon energy selected by a two-crystal Si (111) monochromator. The x-ray microprobe beam was produced using a Fresnel zone plate followed by a 20 μm aperture in a Pt-Ir disk passing only radiation focused to the first order spot (*11,12*). This focusing technique has only recently been made possible by the development of high brilliance synchrotron x-ray sources. The $(0,0,1-\delta)$ SDW reflection was measured with a probe spot size of 0.5 μm using 5.8 keV incident photons. This photon energy is below the 5.95 keV threshold for exciting Cr K shell characteristic fluorescence, eliminating a source of background. To study the $(0,0,2-2\delta)$ CDW reflection at nearly the same diffractometer angles, the photon energy was tuned to 11.6 keV. The maximum count rates from the SDW and CDW reflections were 10 and 2500 counts per second, respectively.

The scattering plane for our diffraction experiments, defined by the wavevectors **k** and **k'** of the incident and diffracted beams, was normal to the [110] direction, so that the momentum transfer **k'-k** was along [001] (Fig. 1D). Previous studies of Cr crystals have found that SDW domains with a component of **Q** along the surface normal are preferentially occupied in the near-surface region due to strain effects (*9,13*). In order to ensure a diversity of SDW domain orientations, our experiments were performed with a polished Cr (111) single crystal (purchased from Alfa Aesar, Inc.) in which we observed all three possible orientations of **Q** using CDW reflections in rotating anode x-ray diffraction experiments.

Images of individual domains of **Q** were formed by measuring the intensity of either the SDW or CDW reflection as a function of the position of the sample under the beam. The equivalence in our geometry of the two reflections as means of observing **Q** domains at temperatures above $T_{SF}$ will be demonstrated below. An image of the (0,0,1-δ) SDW intensity, collected with a 20 μm step size at 130 K is shown (Fig. 2A). Scanning the same area using the (0,0,2-2δ) CDW reflection reveals a nearly identical domain structure (Fig. 2B) where the areas of higher intensity enclosed by black lines correspond to regions in which the propagation wave vector **Q** is along [001]. The roughly 20 μm absorption length for both 5.8 and 11.6 keV x-rays leads to significant contributions to the diffracted signal from depths of up to several microns. The characteristic length scale of the domains ranges from tens to hundreds of microns. In contrast to the domain patterns typical of ferromagnets, the domains of **Q** are highly irregular in shape and size suggesting that relatively weak effects due to inhomogeneity dictate the domain geometry.

The intensity map of the SDW reflection can be understood in the context of the dependence of the magnetic x-ray scattering cross section on the spin polarization **S** and the directions $\hat{\mathbf{k}}$ and $\hat{\mathbf{k}}'$ of the incident and diffracted beams. Two spin terms contribute to the non-resonant cross section for scattering which is proportional to $\left|\mathbf{S}\cdot(\hat{\mathbf{k}}\times\hat{\mathbf{k}}')\right|^2 + \left|\mathbf{S}\cdot\hat{\mathbf{k}}(1-\hat{\mathbf{k}}\cdot\hat{\mathbf{k}}')\right|^2$ (*14*). With our diffraction conditions, the first term dominates when **S** is not in the scattering plane. Because the sample was mounted with [110] normal to the diffraction plane, the magnetic scattering cross sections for the two possible transverse spin polarizations were equal. In the areas of intense scattering in Fig. 2A the SDW domains thus have **Q** along [001] and **S** perpendicular to **Q**. Spins oriented in the scattering plane (with **S** along the [001] direction, i.e. the longitudinal SDW below $T_{SF}$) have a magnetic scattering cross section resulting only from the second term above, which is reduced by a factor of 30 in comparison with the cross section for spins along perpendicular directions. Scans along (0,0,*l*) in reciprocal space taken at points on and off an area of intense scattering (Fig. 2C) illustrate that the contrast in Fig. 2A arises from the SDW domains and not simply from the variation of the position of the SDW peak in reciprocal space across the sample.

We have exploited the dependence of the magnetic scattering cross section on **S** to study the spin flip transition in detail. The intensity of magnetic scattering from domains in which **Q** is in the scattering plane is drastically decreased upon cooling through $T_{SF}$. This effect has been previously observed in x-ray experiments, but without spatial resolution (*9*). Selection rules applicable to neutron scattering can lead to a similar decrease in the scattered neutron intensity below $T_{SF}$ (*8*). The SDW domain marked with a star is shown in images made with a 3 μm step size (Fig. 3). As the sample was cooled through the spin flip transition in a series of steps from 140 K to 110 K, the region in which scattering was observed decreased in extent and disappeared as the spins reoriented to point along **Q**. During each measurement the sample reached a steady-state temperature that was held for several hours – these can thus be considered maps of the magnetic domains in thermal equilibrium. Comparing the image at 130 K with one taken at that temperature upon warming (not shown) reveals that thermal hysteresis is negligible on the micron scale. In contrast to the SDW, images made at 130 K and 110 K using the CDW reflection from the same domain (Fig. 4) illustrate that scattering from the CDW is the same above and below $T_{SF}$. The extent of the **Q** || [001] domain is thus unchanged by the spin flip transition. The dramatic drop in the intensity of magnetic scattering observed in Fig. 3 can be attributed solely to the reorientation of the spins below $T_{SF}$ from transverse to longitudinal upon cooling.

Because the spin flip transition is first order, with a well-known latent heat and change in volume (*7*), it can be expected to take place at a single temperature in a homogeneous system. The fact that different points in the images of Fig. 3 are converted from blue to yellow-green at different temperatures shows that this is not the case for our Cr sample. The variation in transition temperature across the **Q** || [001] domain is illustrated by Fig. 1B, in which the mean intensity in four regions of the images of Fig. 3 is plotted as a function of temperature and set against the magnetization jump which is the bulk signature of the spin flip transition. At each temperature, the positions of the region of integration were adjusted to compensate for linear shifts in the position of the beam on the sample due to contraction of sample holder upon cooling. Although points 3 and 4 are only 5 μm apart, the spin flip transition temperature at the two points differs by 5 K. The count rate in area 1, where $T_{SF}$ is lowest, remains constant as the spin flip transition occurs at points 3 and 4, falling only at a lower temperature. The mean intensity in area 2, a region in which **Q** is perpendicular to [001], and there are neither (0,0,1-δ) SDW nor (0,0,2-2δ) CDW peaks, is unchanged passing through $T_{SF}$.

A phase transition broadened by impurities or other defects is common enough - the defects nucleate the low-temperature phase, which then grows into the regions occupied by the high-temperature phase (*15*). A recent example of this is the growth of ferromagnetism starting at natural and artificial grain boundaries in manganite films (*16*). What we have discovered here is the nucleation of a magnetic phase not at a structural grain boundary, as for the manganite films, but at its magnetic equivalent. Indeed, the combination of our x-ray charge (Fig. 4) and spin (Fig. 3) images of magnetic domains in Cr clearly show that the spin flip transition begins at the boundary between **Q** domains, across which there is a 90° rotation of **Q**, and progresses inwards.

There are three microscopic effects that could be responsible for our discoveries. The first is somewhat similar to what occurs for exchange biasing, a crucial concept underlying useful magnetoresistive devices (*17*) and is based on a local moment model that considers only short-range exchange interactions, as for an insulator. The key insights are that one of the transverse spin directions on one side of a 90° **Q** domain wall actually points along the longitudinal direction on the other side, and that a Heisenberg exchange interaction would favor antiparallel rather than perpendicular alignment. Inevitably, due to disorder, one of the transverse domains will be more stable than the other, allowing it, via the exchange interaction, to nucleate the longitudinal state in the less stable transverse domain. Another mechanism, very much analogous to what is suspected for the manganite films (*16*), invokes strain fields, which may be either the cause or the effect of the **Q** domain walls. The idea here is that $T_{SF}$, is a strong function of strain (*18,19*), and that near the **Q** domain walls, the strain is sufficient to raise $T_{SF}$ by the observed amount. The final and in many respects most interesting candidate mechanism invokes the quantum mixing of states with different Fermi surfaces across the domain wall. The mixing will result in a different spin-orbit coupling near that interface, which could then stabilize the longitudinal phase at a higher temperature.

The magnetic x-ray microscope images described here are an important first step towards understanding the relationship of microscopic domains to the bulk properties of this common element and prototypical magnet.  The interplay between macroscopically observable phenomena and the configuration of domains and domain walls, both magnetic and ferroelectric, is becoming increasingly important as density requirements in many technologies drive device feature sizes towards those of individual domains.  Recent achievements of multilayer magnetic devices with antiferromagnetic coupling layers have stimulated a renewed interest in chromium and other antiferromagnetic materials (*20-22*).  Our experiments are significant beyond the world of Cr because the ability to image antiferromagnetic domains with the same ease with which magnetic force microscopy can image ferromagnetic domains is a prerequisite for the scientifically-based engineering of antiferromagnets.

**Fig. 1** (**A**) At the spin flip transition temperature $T_{SF}$, the spin polarization **S** changes from transverse to longitudinal with respect to the modulation wavevector **Q**. The period of the spin modulation is incommensurate with the lattice spacing *a*. (**B**) (upper panel) A step in the bulk magnetization of Cr in an applied field of 2 kG serves as a marker for the spin flip transition. (lower panel) The mean intensity in the numbered areas of Fig. 3 plotted as function of temperature demonstrates that $T_{SF}$ varies by more than 7 K across this single SDW domain. (**C**), The positions of the (002) allowed Bragg reflection (solid cube), charge density wave satellite (open cubes), and spin density wave satellite (spheres) peaks are shown in the regions of reciprocal space near (001) and (002). (**D**) The incident and diffracted beams **k** and **k'** for a **Q** ∥ [001] domain near a **Q** domain wall. Transverse-polarized spins in this domain (dashed lines, labeled $S_T$) have components not in the diffraction plane and thus result in more intense scattering than spins polarized along **Q** (solid line, $S_L$). Ordered spins in the neighboring domain with **Q** ∥ [100] result in no diffracted intensity because the Bragg diffraction condition is not satisfied.

**Fig. 2** Magnetic domains in Cr at T=130 K. Maps of the intensity of the (**A**), spin density wave reflection at (0,0,1-δ) and (**B**), charge density wave reflection at (0,0,2-2δ) in a 500 μm × 500 μm region. These images (including those in Figs. 3 and 4) have been resampled on a grid with four times the point density of the original scan and smoothed with a three pixel square moving average. The intensity scale for these images ranges from yellow (lowest count rate) to blue (highest count rate). The saturation level and threshold for the domain outlines are chosen to emphasize the domain contrast and are scaled between the two images by the ratio of the mean count rate. (**C**) Reciprocal space scans along $(0,0,l)$ at the positions indicated on the spin density wave map. The solid line is a Gaussian fit to the scan at position 1.

**Fig. 3** Images of a single spin density wave domain at temperatures near the spin flip transition temperature. This domain also appears at a different scale in Fig. 2A where it is marked with a star. The transition from transverse to longitudinal spin polarization at $T_{SF}$ results in the disappearance of magnetic scattering from the SDW domain.

**Fig. 4** Images of the charge density wave reflection intensity in the area of Fig. 3 at (**A**) 130 K and (**B**) 110 K. The Q orientation and size of the SDW domain are unchanged by cooling through $T_{SF}$.

Evans *et al.*, Figure 1

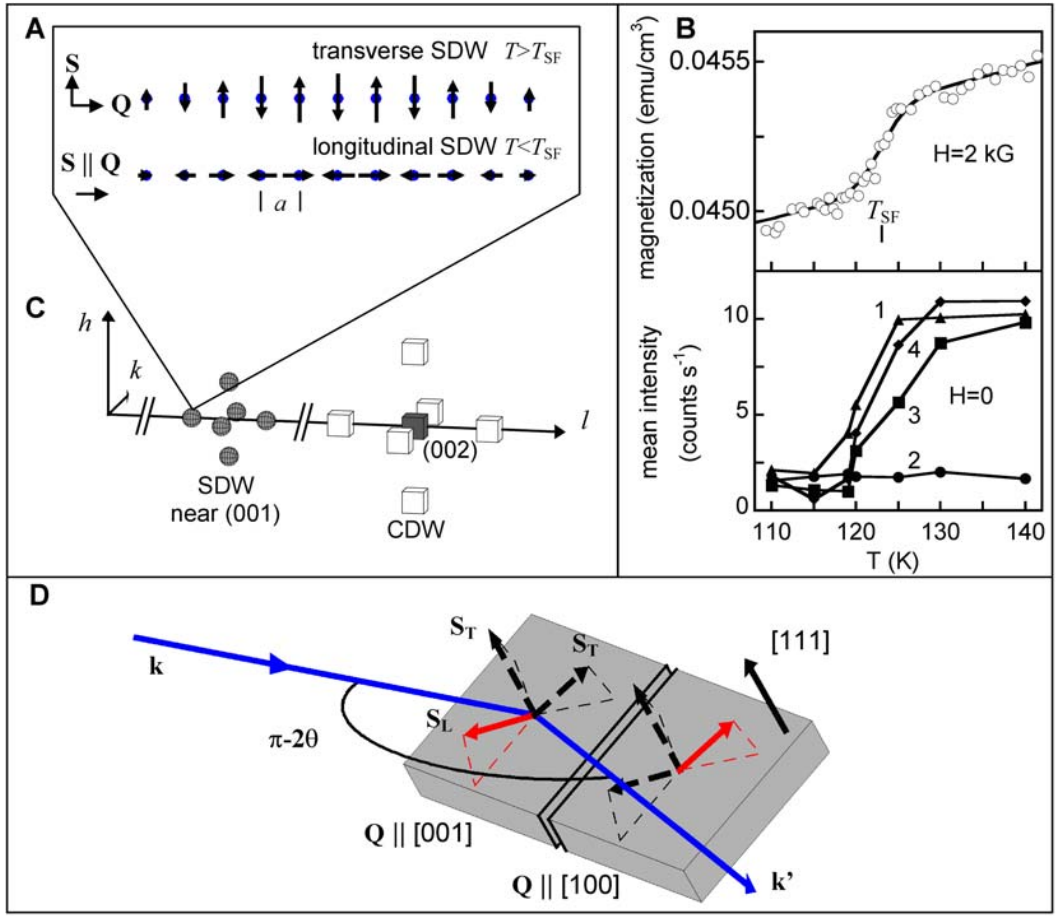

Evans *et al.*, Figure 2

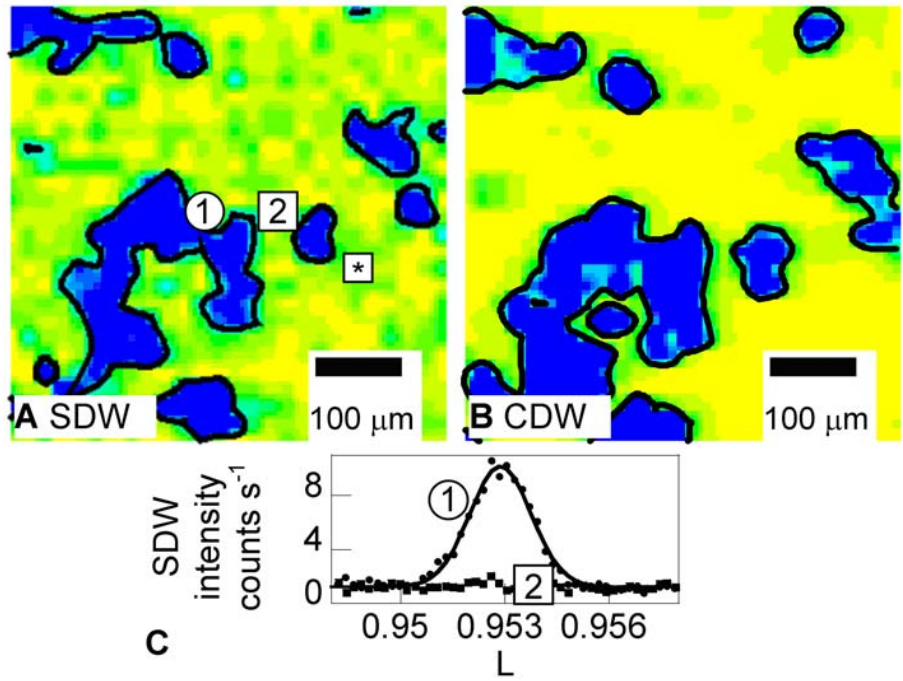

Evans et al., Figure 3

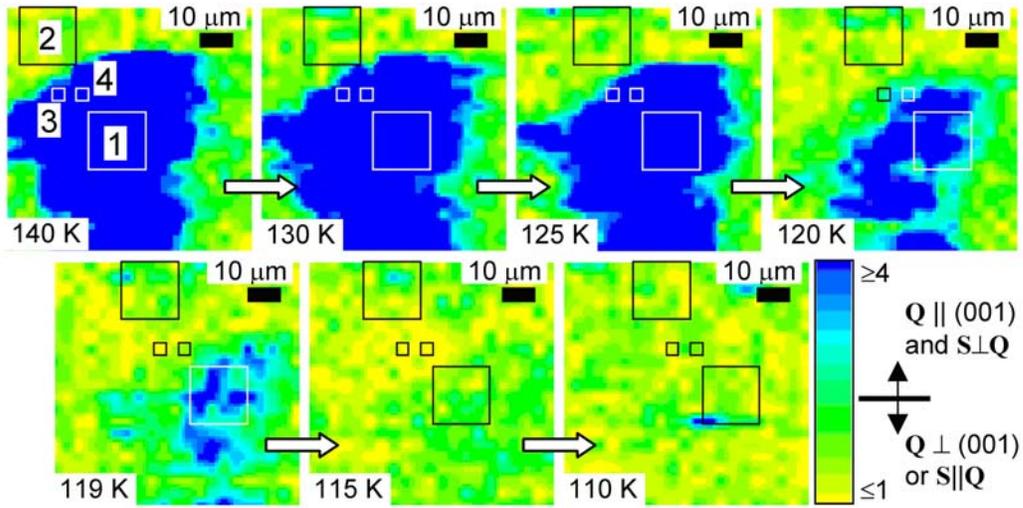

Evans *et al.*, Figure 4

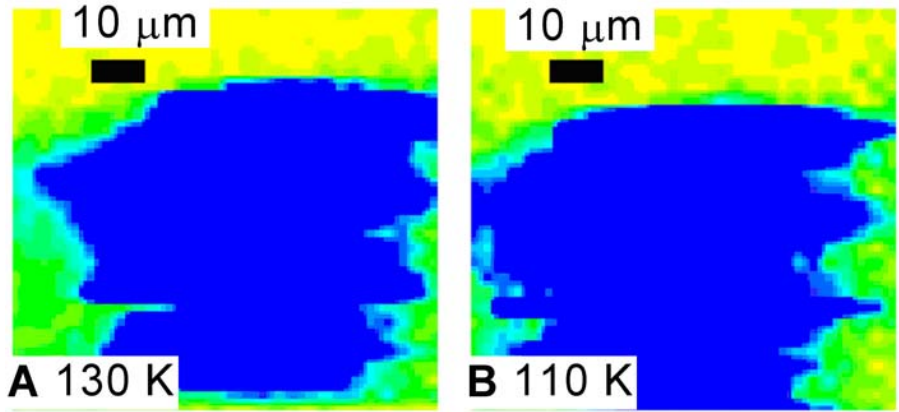